\documentclass[twoside]{article}
\usepackage{fleqn,espcrc2}

\usepackage[dvips]{graphicx, color}

\newcommand{\ncm}{\newcommand}
\newcommand{\rencm}{\renewcommand}
 
\def\a{\alpha}

\def\f{\varphi}    

\def\l{\lambda}
\def\m{\mu}

\def\o{\omega}
\def\p{\pi}       
\def\r{\rho}      

\def\D{\Delta}

\def\fh{\hat{\f}}

\ncm{\dsp}{\displaystyle}
\ncm{\nn}{\nonumber}
\ncm{\nnn}{\nonumber\linebreak[4]}
\ncm{\nit}{\noindent}
\ncm{\del}{\partial}
\ncm{\av}[1]{\mbox{$\langle #1 \rangle$}}
\ncm{\avc}[1]{\mbox{$\langle #1 \rangle_{\psi}$}}
\ncm{\half}{\mbox{{\small $\frac{1}{2}$}} }
\ncm{\quart}{\mbox{{\small $\frac{1}{4}$}} }
\ncm{\tq}{\mbox{{\small $\frac{3}{4}$}} }
\ncm{\third}{\mbox{{\small $\frac{1}{3}$}} }
\ncm{\sixth}{\mbox{{\small $\frac{1}{6}$}} }
\ncm{\eigth}{\mbox{{\small $\frac{1}{8}$}} }
\ncm{\thrhalf}{\mbox{{\small $\frac{3}{2}$}} }
\ncm{\thrfor}{\mbox{{\small $\frac{3}{4}$}} }
\ncm{\twothi}{\mbox{{\small $\frac{2}{3}$}} }
\ncm{\fivtwo}{\mbox{{\small $\frac{5}{2}$}} }

\ncm{\dxx}{\mbox{$\partial_{x}^2$}}
\ncm{\dx}{\mbox{$\partial_{x}$}}
\ncm{\dt}{\mbox{$\partial_{t}$}}
\ncm{\dtt}{\mbox{$\partial_{t}^2$}}
\ncm{\un}{1\!\!1}
\ncm{\RE}{\mbox{Re}}
\ncm{\IM}{\mbox{Im}}
\ncm{\Tr}{\mbox{tr}\,}
\ncm{\diag}{\mbox{diag}\,}
\ncm{\Det}{\mbox{Det}\,}
\ncm{\Log}{\mbox{Log}\,}
\ncm{\ra}{\rightarrow}
\ncm{\la}{\leftarrow}
\ncm{\dg}{\dagger}
\ncm{\pr}{\prime}
\ncm{\ha}{\hat{a}}
\ncm{\hP}{\hat{P}}

\ncm{\aplt}{ \mbox{}_{\textstyle \sim}^{\textstyle < }     }
\ncm{\apgt}{ \mbox{}_{\textstyle \sim}^{\textstyle > }     }
\ncm{\Oa}{\mbox{$\mbox{O}(a)$}}
\ncm{\Sp}{\mbox{\hspace{1.0cm}}}
\ncm{\capit}[1]{\caption{\it #1}}

\def\be{\begin{equation}}
\def\ee{\end{equation}}
\def\bea{\begin{eqnarray}}
\def\eea{\end{eqnarray}}
\def\bi{\begin{itemize} \itemsep = 0.01\itemsep  }
\def\bii{\begin{itemize} \small \itemsep = 0.01\itemsep }
\def\ei{\end{itemize}}
\def\bc{\begin{center}}
\def\ec{\end{center}}
\def\bs{\begin{slide}}
\def\es{\end{slide}}

\def\beac{\begin{eqnarray} \color [rgb] {0,0,1} }
\def\eeac{\end{eqnarray} }

\ncm{\shead}[1]{\bc { \Large \color [rgb]{1.0, .0, .0} #1 \normalcolor} \ec}
\ncm{\ssubh}[1]{{\large \color [rgb]{1.0,.0,.1} #1 \normalcolor}}

\rencm{\thefootnote}{\mbox{\protect{$\fnsymbol{footnote}$}} }

\ncm{\front}[5]   
{
   \begin{titlepage}
      \noindent {#1} \hfill {#2}\\
      \begin{center}
         \vspace{1.5\baselineskip}
         {\Large\bf  #3  } \\
         \vspace{2\baselineskip}
         \vspace{1.5\baselineskip}
          #4\\
         \vspace{1.5\baselineskip}
   
         Institute of Theoretical Physics, \\
         Valckenierstraat 65, 1018 XE Amsterdam,
         The~Netherlands.
    
      \end{center}
      \vfill
      {\bf Abstract}\\
       #5
   \end{titlepage} 
}

\ncm{\frontslide}[4]
{
   \begin{titlepage}
      \noindent {#1} \hfill {#2}\\
      \begin{center}
         \vspace{1.5\baselineskip}
         {\Large\bf  #3  } \\
         \vspace{2\baselineskip}
         \vspace{1.5\baselineskip}
          #4\\
         \vspace{1.5\baselineskip}
   
         Institute of Theoretical Physics, \\
         Valckenierstraat 65, 1018 XE Amsterdam,
         The~Netherlands.
    
      \end{center}
   \end{titlepage} 
}

\begin{document}

\title{Finite Temperature Simulations from Quantum Field Dynamics?
}

 \author{ Mischa~Sall\'e, Jan~Smit and Jeroen C.~Vink\thanks{ 
 Presented by J.~Vink 
 } \\[0.2cm]
 Institute for Theoretical Physics, University of Amsterdam\\
 Valckenierstraat 65, 1018 XE Amsterdam, The Netherlands
}



\begin{abstract}
We describe a Hartree ensemble method to approximately solve the 
Heisenberg equations for the $\f^4$ model in $1+1$ dimensions. 
We compute the energies and number densities of the
quantum particles described by the $\f$ field and
find that the particles initially thermalize with a Bose-Einstein 
distribution for the particle density.  Gradually, 
however, the distribution changes towards classical equipartition. 
Using suitable initial conditions quantum thermalization is achieved much 
faster than the onset of this undesirable equipartition. We also show how 
the numerical efficiency of our method can be significantly improved.
\vspace{1pc}
\end{abstract}

\maketitle

\section{Hartree ensemble approximation}

Non-perturbative equilibrium properties of quantum field theories are studied with 
great success using numerical lattice techniques.
Similarly one would like to investigate
quantum field {\em dynamics} non-perturbatively. This has obvious applications
to e.g.\ heavy ion collisions and early universe physics. Furthermore,
it could be used to study equilibrium properties in cases where 
Monte Carlo methods do not work, such as in QCD at
finite density or chiral gauge theories.
Using Monte Carlo techniques to compute the Minkowski time path integral is, 
of course, not possible. 
Neither can one directly solve the Heisenberg equations for the
quantum field operators.
Hence one has to resort to approximations, such as
classical dynamics \cite{GrRu88} (for recent work see \cite{AaBo00}),
large $n$ or Hartree (see e.g.~\cite{CoHa98}). 
Here we discuss a different type
of Hartree or gaussian approximation than used previously, in which 
we use an ensemble of gaussian wavefunctions to compute Green functions.
Ref.~\cite{SaSmVi00} also contains a discussion of this method and
a detailed presentation will appear elsewhere.

We use the lattice $\f^4$ model in $1+1$ dimensions as a test model, which has
the following Heisenberg operator equations,
\be
\dot{\hat{\f}} = \hat{ \p}, \;\;\;
\dot{\hat{\p}} = (\D - \m^2)\hat{\f} - \l \hat{\f}^3, \label{eq:heis}
\ee
with $\D$ the lattice laplacian, $\m$ the bare mass parameter 
and $\l$ the coupling constant (we use lattice units $a=1$).
The Hartree approximation assumes that the wavefunction
used to compute the Green functions is of gaussian form, such that
the field operator can be written as
\be
\fh_x  =  \f_x + \sum_\a[ f^\a_x \hat{a}_\a + f^{* \a}_x \hat{a}^{\dagger}_\a],
\ee
with $\hat{a}^{\dagger}_\a$ and $\hat{a}_\a$ time-independent creation and
annihilation operators and time-dependent mean field $\f$ and mode functions $f^\a$.
All information is contained in the one- and two-point functions, which 
can be expressed (for gaussian pure states) as
\be
 \av{ \fh_x } = \f_x,\;\;
 \av{ \fh_x \fh_y }_{\rm conn} =  \sum_\a f_x^\a f_y^{\a *}.
\ee
The mode functions represent the width of the wave function, allowing
for quantum fluctuations around the mean field. 
With Hartree dynamics we therefore expect to improve the classical dynamics.
Of course we cannot expect to capture all quantum effects, e.g.
tunneling is beyond the scope of the gaussian approximation. 

The Heisenberg equations (\ref{eq:heis}) provide self-consistent
equations for the mean field $\f$ and the mode functions $f^\a$,
\bea
   \ddot{\f}_x &\!\!\! =\!\!\! & \D\f_x - [ \m^2 + \l (\f_x^2 +
             3\sum_{\a} f_x^{\a} f_x^{\a *}) ]\f_x \nonumber \\
   \ddot{f}_x^{\a} &\!\!\! = \!\!\!& \D f_x^\a - [ \m^2 + \l (3\f_x^2 +
      3\sum_{\a} f_x^{\a} f_x^{\a *}) ] f_x^\a  \label{eq:eom}
\eea

In order to simulate a more general density matrix instead of the pure
gaussian state assumed in eq.~(\ref{eq:eom}), we average
over a suitable ensemble of Hartree realizations by specifying different
initial conditions and/or coarsening in time.
In this way we compute Green functions with a non-gaussian density matrix 
$\hat\r = \sum_i p_i \hat\r_i^G$, as
\bea
  \av{ \hat\f_x \hat\f_y }_{\rm conn} &=&  
   \sum_i p_i [  \av{ \hat\f_x \hat\f_y }_{\rm conn}^{(i)} 
                      + \f_x^{(i)} \f_y^{(i)}]  \nonumber \\
   & - & (\sum_i p_i \f_x^{(i)})(\sum_j p_j \f_y^{(j)}).
\label{eq:avgx}
\eea

It should be stressed that for typical Hartree realizations the mean field 
$\f_x$ is {\it inhomogeneous} in space. This is different from the ensemble average
of $\fh$ which may, of course, be homogeneous.
We also note that the equations (\ref{eq:eom}) 
can in fact be derived from a hamiltonian.
Since the equations are also strongly non-linear, this suggests that the
system will evolve to an equilibrium distribution with equipartition of energy,
as in classical statistical physics.

\section{Observables}
To assess the viability of our Hartree ensemble approximation, we
solve the equations (\ref{eq:eom}) starting from a number of initial conditions. 
We compute the particle number densities and energies from the 
connected two-point functions of $\f$ and $\p$, after coarse-graining in space 
and time and averaging over initial conditions, 
\bea
\frac{1}{N}\sum_{xz}e^{-ikz} \av{\overline{ \hat\f_{x+z} \hat\f_x} }_{\rm conn}
    & = & \frac{n_k + \half}{\o_k}, \\
\frac{1}{N}\sum_{xz}e^{-ikz} \av{ \overline{ \hat\p_{x+z} \hat\p_x} }_{\rm conn}
   &  =  & (n_k + \half) \o_k.
\label{eq:nando}
\eea
The over-bar indicates averaging over some time-interval and initial conditions
as in eq. (\ref{eq:avgx}),
$\o_k$ is the energy, $n_k$ the (number) density of particles with
momentum $k$ and $N$ the number of lattice sites. 

For weak couplings, such as we will use
in our numerical work, the particle densities should have a Bose-Einstein (BE)
distribution,
\be
  n_k = 1/(e^{\o_k/T} - 1),\, \o_k^2 = m^2(T) + 2 - 2\cos(k),
\label{eq:freeform}
\ee
with $T$ the temperature and $m(T)$ an effective  finite temperature mass.
We have verified that this is indeed the case using Monte Carlo simulations of 
the euclidian time version of the model.

\begin{figure}[t]
\hspace{0.0cm}
\vspace{-.3cm}

\scalebox{0.30}[0.30]{ \rotatebox{270}{\includegraphics[clip]{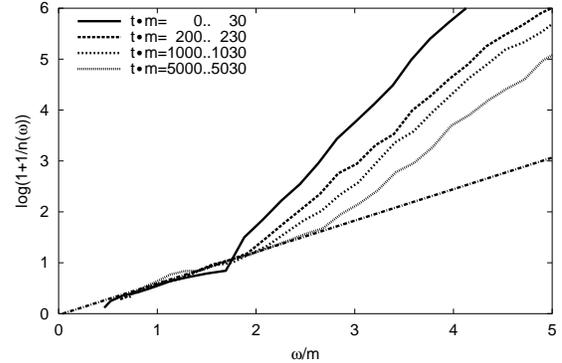} } }
\vspace{-1.0cm}
\caption{
Particle number densities as a function of $\o$ at various times. 
The straight line gives a BE distribution with temperature
$T/m = 1.7$. (The lattice volume $Lm=32$, coupling $\l/m^2=0.083$ 
and inverse lattice spacing $1/am = 8$; $m \equiv m(T=0)$).
\vspace{-0.7cm}
\label{fig:pa} }
\end{figure}

\section{Numerical results}
First we use initial conditions which correspond to fields far out of
equilibrium: gaussians with mean fields that have just
a few low momenta modes,
\be
\f_x = v,\; \p_x = A \sum_j^{j_{\rm max}}   \cos(k_j x + \a_j).
\ee
The $\a_j$ are random phases and the $A$ is a suitable amplitude.
Initially the mode functions are plane waves, $e^{ik x}/\sqrt{2\o L}$,
such that all energy resides in the mean field.

The results in Fig.~\ref{fig:pa} show that fast, well before $tm\approx 200$, 
a BE distribution
is established for particles with low momenta. Slowly this thermalization
progresses to particles with higher momenta, while the temperature 
$T \approx 1.7m$ remains roughly constant. Such a thermalization does
{\em not} happen when using a single Hartree realization with a {\em homogeneous}
 mean field $\f$. The difference may be understood, because
particles in our method can scatter via the inhomogeneities in the mean field.

\begin{figure}[t]
\hspace{-0.6cm}
\scalebox{0.88}[0.72]{ \includegraphics[clip]{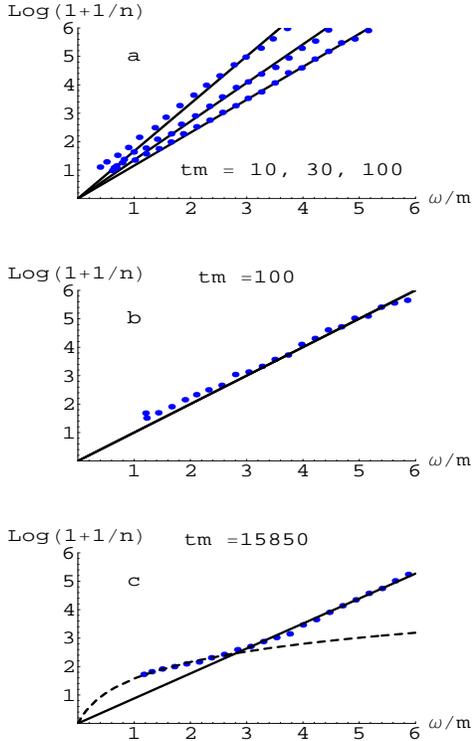} } \hspace{-0.7cm}
\vspace{-0.9cm}
\caption{ Particle number densities as a function of $\o$, at early times
(a and b) and large time (c).
 ($Lm=25.6$, $\l/m^2=0.5$ and $1/am = 10$).
 }
\label{fig:combi}
\vspace{-0.7cm}
\end{figure}

Next we want to speed-up the thermalization of the high momentum modes.
Therefore we use different initial conditions in which the energy
is distributed more realistically over the Fourier modes of the mean
field. The modes are the same as before but mean fields are drawn from
an ensemble with a BE-like probability distribution,
\be
 p(\f_k, \p_k) \! \propto \!  \mbox{exp}
          [ -(e^{\o_k/T_0}-1)( \p_k^2  + \o_k^2 \f_k^2)/2\o_k ].
\ee
We also want to probe large time scales, which were not yet reached in the
weak coupling simulation of fig.~\ref{fig:pa}.  

Fig.~\ref{fig:combi}a shows the particle density computed {\it without} 
the mean field part in eq.~(\ref{eq:avgx}). Already at $tm=20$ a credible
BE distribution has established, with a temperature that increases 
until it saturates at $T \approx 1.0m$ at $tm=100$ ($T=0$ at $t=0$). 
Fig.~\ref{fig:combi}b shows the densities at $tm=100$, 
now with the mean field contribution included. The straight line is a BE
distribution at $T\approx 1.01m$.

At very large times, Fig.~\ref{fig:combi}c, the BE distribution 
persists in the large $\o$ region,  albeit with a slightly higher 
temperature $T \approx 1.14m$, but for small $\o$ one clearly sees deviations. 
The particle densities
here can be fitted rather well with the classical form $n = T_{cl}/\o$ 
(dashed curve). The corresponding classical temperature slowly decreases, 
as energy
disperses more and more to the large momentum modes. It is remarkable that
the ``decay time'' of this temperature is some two orders of magnitude larger 
than the warming-up time shown in Fig.~\ref{fig:combi}a. 

\begin{figure}[t]
\scalebox{0.80}[0.80]{ \includegraphics[clip]{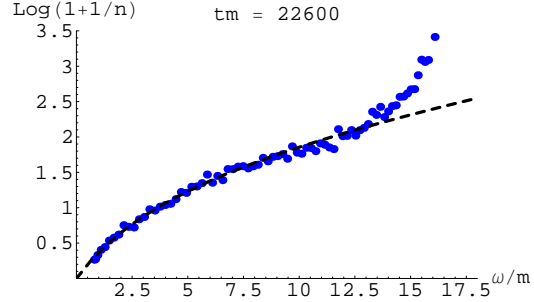} }
\vspace{-1.5cm}
\caption{ Particle number densities at large
times, $tm = 25600$ and high initial temperature, $T_0=5m$. 
 ($Lm=25.6$, $\l/m^2=0.5$ and $1/am = 10$).
 }
\label{fig:late}
\vspace{-0.7cm}
\end{figure}

Finally, in Fig.~\ref{fig:late}, we
show data for a large time $tm \approx 25000$ and at a higher (initial)
temperature, $T_0=5m$. 
Here equipartition has extended to all modes with $\o \aplt 12m$,
as is evident from the accuracy of the fitted curve
$n=1.9m/\o$ (dashed line).

\begin{figure}[t]
\hspace{-0.6cm}
\scalebox{0.74}[0.75]{ \includegraphics[clip]{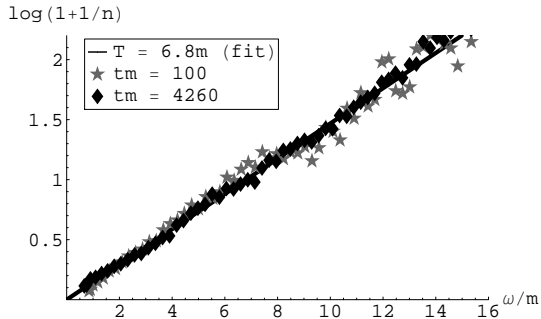} } \hspace{-0.7cm}
\vspace{-1.0cm}
\caption{ Particle number densities as a function of $\o$,
starting from BE type initial conditions.
The straight line through the origin is a BE distribution with
temperature $T/m=6.7$. ($Lm=23$, $\l/m^2=0.083$ and $1/am = 22$). }
\vspace{-0.9cm}
\label{fig:be}
\end{figure}

Even though the timescales of relaxation to a BE distribution and relaxation
to classical equipartition appear to be widely separated, the simulations
of Fig.~\ref{fig:combi} could not easily be used to study finite temperature
equilibrium physics: only around $tm=100-200$ there may be a window in which
the fields have thermalized to a BE distribution, without significant deviations
due to the up coming classical equipartition. To increase such a window
we did a simulation at weaker coupling and higher temperature.
Now we find density distributions as shown in Fig.~\ref{fig:be}.
Already after a relatively short time, before $tm\approx 100$, the particles 
have acquired a BE distribution up to large momenta $\o/m\approx 12$.
Note that even with the BE-type initial conditions, the fields initially are
out of equilibrium: energy is initially carried by the mean field only but 
within a time span of $tm\approx 200$ it is redistributed over the modes.

\begin{figure}[t]
\scalebox{0.75}[0.61]{ \includegraphics[clip]{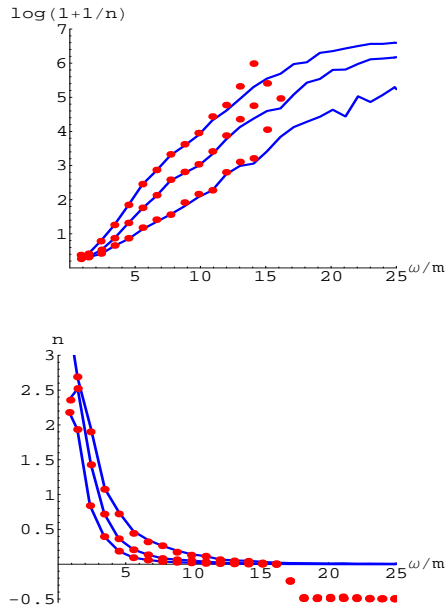} }
\vspace{-0.9cm}
\caption{ Hartree dynamics with all mode functions (drawn lines)
and with a reduced number (32 out of 128) mode functions (dots).
($Lm=5.7$, $\l/m^2=0.083$ and $1/am = 22$).
\label{fig:compare}}
\vspace{-0.7cm}
\end{figure}

Finally we try to improve the efficiency of our method. Since we have
to solve for $N$ mode functions, on a lattice with $N$ sites, the CPU
time for one time-step grows $\propto N^2$. However, mode functions 
corresponding
to particles with momenta much larger than $T$ should be irrelevant,
since these particle densities are exponentially suppressed. This suggests
that we can discard such modes. This is tested in Fig.~\ref{fig:compare},
where we compare simulation results using all mode functions with results
obtained using only a quarter of the mode functions. This corresponds with
mode functions with initial plane wave energy $\o \aplt 17m$. 
Clearly the results
for particles with significant densities $n$ are indistinguishable.
For particles with energy larger than $\approx 17m$, there are no longer
mode functions that can provide the vacuum fluctuations and consequently
the particle density defined by (\ref{eq:nando}) drops to $-1/2$.

\section{Conclusion}
Using a Hartree ensemble method, we have demonstrated that
we can simulate approximate quantum thermalization in a weakly coupled scalar 
field model in real time. 
Only after times much longer than typical equilibration 
 times,  the approximate nature of the dynamics shows up in deviations
from the BE distribution towards classical equipartition. 
We have furthermore found that these simulations can be done efficiently 
using a limited number of mode functions.

\end{document}